\documentclass[runningheads]{llncs}
\usepackage[T1]{fontenc}
\usepackage{tablefootnote}
\usepackage{graphicx}
\usepackage{adjustbox}
\usepackage{makecell}

\begin{document}
%
\title{Who ruins the game?: unveiling cheating players in the ``Battlefield" game}
\titlerunning{Who ruins the game?: unveiling cheating players in the ``Battlefield" game}





\author{Dong Young Kim \and Huy Kang Kim}
\institute{School of Cybersecurity, Korea University, Republic of Korea \\ 
\email{\{klgh1256, cenda\}@korea.ac.kr}}
\authorrunning{D. Y. Kim and H. K. Kim}
\maketitle              
\begin{abstract}
The ``Battlefield" online game is well-known for its large-scale multiplayer capabilities and unique gaming features, including various vehicle controls. However, these features make the game a major target for cheating, significantly detracting from the gaming experience.

This study analyzes user behavior in cheating play in the popular online game, the ``Battlefield", using statistical methods. We aim to provide comprehensive insights into cheating players through an extensive analysis of over 44,000 reported cheating incidents collected via the ``Gametools API". Our methodology includes detailed statistical analyses such as calculating basic statistics of key variables, correlation analysis, and visualizations using histograms, box plots, and scatter plots. Our findings emphasize the importance of adaptive, data-driven approaches to prevent cheating plays in online games.

\end{abstract}
%
%
%


\section{Introduction}
The Battlefield game~\cite{battlefieldgames} (operated by Electronic Arts Inc.) is a large-scale team multiplayer online game. It is also well-known for its unique gaming features, such as enabling various vehicle controls (\textit{e.g.}, Tank, Armored Personnel Carrier, Infantry Fighting Vehicle, Fighter Jet, and Attack Helicopter) by game players. Such an aspect distinguishes it from other First-Person Shooters (FPS) games, attracting gamers worldwide. As the same as the real-world battlefield, the vehicle's ability is too strong enough to eliminate the opponent player's infantry troops. To leverage vehicles' ability, some players use game cheating programs that are regarded as unfair gameplay. 

\begin{figure}[h]
    \centering
    \includegraphics[width=0.9\linewidth]{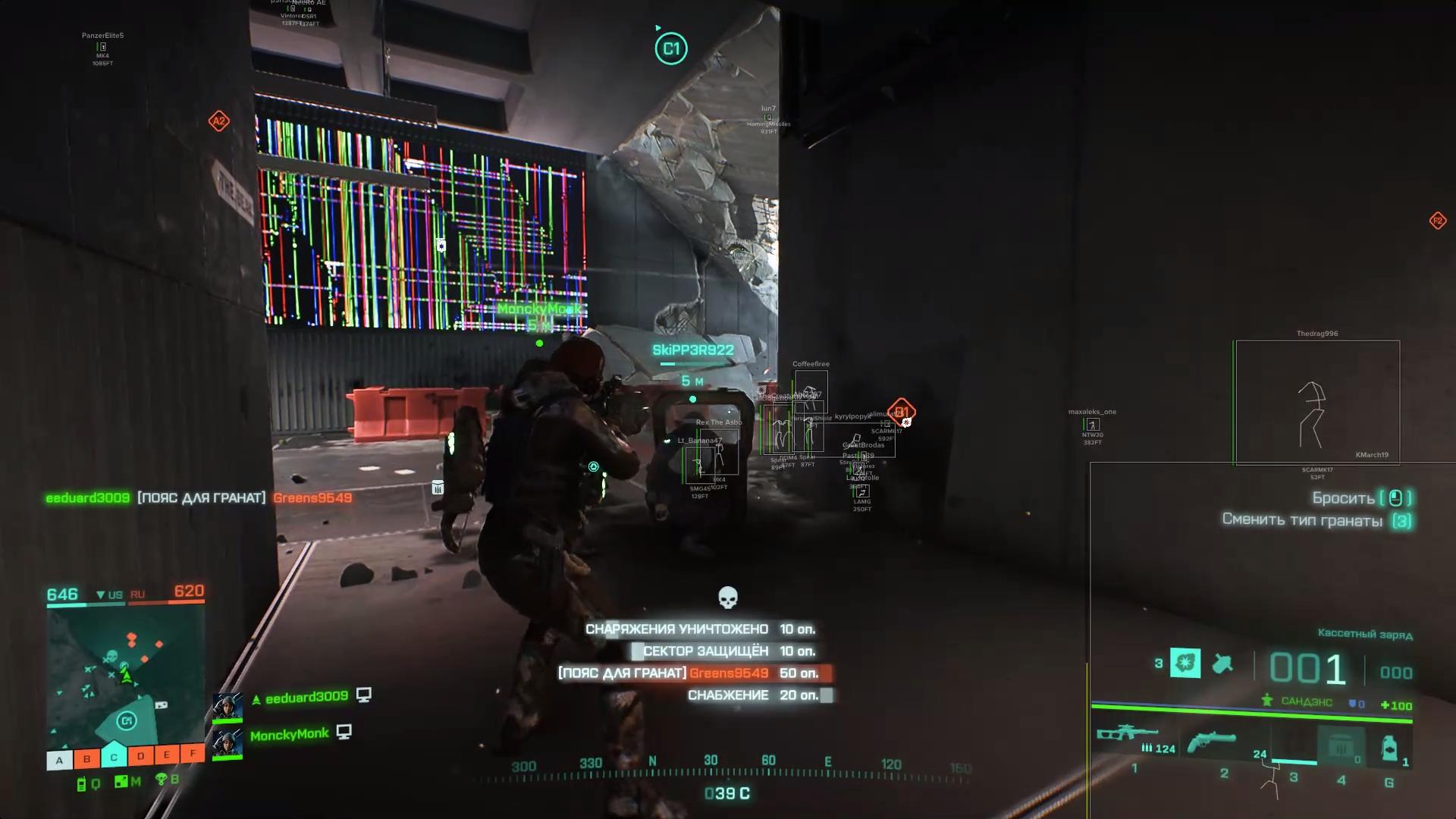}
    \caption{The BattleField in-game cheating play example\protect\footnote{~\cite{battlefieldgamesimg}}}
    \label{fig:the_battleField_game}
\end{figure}

Cheating play is an axis of evil in online gaming. If game companies fail to detect cheating play in time, the game users quit and leave the game because they can feel unfair. Consequently, cheating play detection is a critical success factor in preventing a game company's financial damage and managing the user's loyalty. 

Not only in the Battlefield game, but cheating programs such as Aimbot, Wallhack, Speedhack, Weaponhack, and EXP hack have also been widely used in FPS games for several decades~\cite{han2015online}. 


Thus, like the other FPS games, the Battlefield game has many users complaining about the cheating play. Fig.~\ref{fig:the_battleField_game} shows a cheating player use the Wallhack and ESP cheating tool during the battle. In response, the latest version, the Battlefield 2042, deploys an Anti-Cheat program on the client-side. However, detecting cheating on the client-side may diminish user experience due to program errors and vulnerabilities. Detecting cheating over the network is practically impossible, as it requires real-time monitoring of all users' communications, which could degrade the overall gaming experience due to the real-time nature of FPS games. Server-side detection demands significant resources from game servers. It is practically impossible for game companies to monitor all users' logs, which could increase the game's latency, which is critical in FPS games. Thus, the existing cheating detection methods used in other game genres (\textit{e.g.}, frequently used data mining or machine learning methods in MMORPG genre~\cite{han2022cheating}.) are not easily applicable in FPS games.

Therefore, this paper aims to provide insights for discovering users who use cheating tools without consuming resources from the game servers and clients, based on statistical analysis and visualization.

Our research method employs fundamental statistical analysis, which can effectively distinguish cheating players' behaviors. 
The main contributions of this study can be summarized as follows. 

\begin{itemize}
    \item Unlike existing cheating detection methods in FPS games, our approach does not need to be deployed in-game client-side. The current countermeasures deployed in the game client-side can harm the overall gaming experience because of the overhead (\textit{i.e.}, typical usability vs. security problem)
    \item To the best of our knowledge, it is the first approach for cheating play user detection by lightweight statistical analysis of kill-log in the Battlefield game. We discovered features that can  distinguish cheating players' behaviors successfully.
\end{itemize}

\section{Background}
\subsection{How to play the Battlefield Games?}

  The Battlefield game provides specialized game modes. For instance, some modes focus on the number of infantry kills to win, while others modes focus on the score based on the result of capturing specific areas during the combat. Among various modes, one of the most representative modes is the `Conquest mode'. Conquest mode involves 64 or 128 players, where participants are divided into several classes, such as Assault, Support, Ammunition, and Recon. Each class has specialized equipment. For example, the Support class users, one of the infantry solder classes in the game, carry a medical kit to heal their wounded teammates, or they carry a defibrillator to revive soldiers who have been downed. 

Depending on the map size and scale, these soldiers can control various vehicles. These vehicles include Tanks, Armored Personnel Carriers, Infantry Fighting Vehicles, Anti-Aircraft Vehicles, Fighter Jets, and Attack Helicopters. There are land, water, and air vehicles that players can use from the initial starting point or specific points on the map. The amount of damage dealt by the weapons on these vehicles is significantly higher than that of infantry weapons. Consequently, vehicles can massacre infantry, or infantry may find themselves hiding from the vehicles.

Conquest mode involves infantry using vehicles to capture specific points. Capturing a point occurs when there are more allied soldiers than enemies at a location for a certain period, and it happens faster with more allies present.


The Conquest mode is the most preferable play mode, but cheating play exists in all modes, not only the Conquest mode. 

\subsection{Definition and Types of Cheating in the Game}

The well-known cheating plays in FPS are Aimbot, Damage Modify, Gadget Modify, Wallhack, Stealth, and Magical Bullet. `Aimbot' is a program that artificially enhances a player's shooting accuracy by automatically targeting opponents' critical areas (\textit{e.g.}, headshot), enabling precise shooting. This type of cheating elevates a player's shooting abilities, and eventually disrupting the game's balance seriously. `Damage Modify' is cheating that manipulates a player's weapon to cause more damage than usual, allowing players to overpower enemies with fewer bullets. `Gadget Modify' alters or enhances the functions of the equipment or gadgets players use. For example, it could increase the blast radius of explosives or reduce the cooldown period of gadgets. `Wallhack' makes walls or other obstacles transparent or allows players to see the outlines of enemies through them. It provides a significant tactical advantage by predicting the location and movements of concealed enemies. `Stealth' renders players invisible to enemy sights or detection systems, granting the ability to move covertly and disrupt game balance. `Magical Bullet' includes features that allow players' bullets to penetrate obstacles or automatically target enemies at unrealistically long distances, ignoring ballistics and physical laws, resulting in highly unfair gameplay.

\subsection{Why Cheating Detection in FPS is Important?}
Why is it necessary to prevent cheating in FPS games? First, it is essential to maintain fairness. Ensuring the game's fairness allows all players to compete under equal conditions, making the gaming experience more enjoyable and satisfying. Second, it is to ensure that all users can enjoy the game. One of the main purpose of playing a game is to have fun and feel a sense of accomplishment through challenge. Fun and challenge are key rewards in a game system. If the fun and challenge based reward system is disrubted by some cheating players, then many normal players will quit and leave the game. Third, it disturbs the growth and development of players. Honest players can develop their skills and learn strategies, whereas cheating players skip this developmental process, negatively impacting other players (\textit{i.e.},
contagion)~\cite{woo2018contagion}. Thus, if cheating players gain an advantage in the game, honest players are disadvantaged, leading to many players leaving the game. For these reasons, cheating can ruin the game's overall fair-play culture and lead to user churn.

\section{Related Work}
Tekofsky \textit{et al.}~\cite{tekofsky2015past} analyzed play styles by age in Battlefield 3. Their study used data from 10,416 users to demonstrate that performance and speed of play decrease with age. May \textit{et al.}~\cite{may2024thinking} analyzed how the game Battlefield 2042 implements climate crises and environmental challenges. This study investigated how the game's settings and scenarios encourage players to think about ecological issues and reviewed how user-generated content and discussions about the game promote ecological engagement. They demonstrated through data analysis from 13,129 users how these elements stimulate ecological thinking among players.

Various studies have been conducted on detecting cheating in FPS games. Most of these studies use computer vision to detect changes in status or visually apparent cheating patterns. For example, Nie \textit{et al.}~\cite{nie2024vadnet} identified cheating patterns such as Aimbots in FPS games using a deep learning vision framework called VADNet. Jonnalagadda \textit{et al.}~\cite{jonnalagadda2021robust} proposed a new vision-based approach that captures the final game screen state to detect illegal overlays. They collected data from two FPS games and several cheating software using deep learning models to verify cheat detection performance. Liu \textit{et al.}~\cite{liu2017detecting} detected Aimbots using cosine similarity and statistical methods. Cosine similarity was used to measure the change in aim when players first spot an enemy, calculating the angle between vectors of the player's aiming direction to compare the automatic aiming patterns of Aimbots with those of regular players. Pinto \textit{et al.}~\cite{pinto2021deep} proposed a novel approach using deep learning and multivariate time-series analysis to detect cheats in video games. Instead of game data, they analyzed user behavior data such as key inputs and mouse movements, demonstrating high accuracy in cheat detection. Han \textit{et al.}~\cite{han2015online} discovered several features (\textit{e.g.}, win-rate, headshot-ratio, play counts and play time, etc.) and detection rules to differentiate cheating players in Point Blank, one of the famous FPS games in Asian game market. In this work, the proposed method detected cheating plays based on the analysis of action logs, while our proposed model is based on the analysis of kill-log and stat data.

\section{Methodology}

\vspace{-1cm}

\begin{figure}[!h]
    \centering
    \includegraphics[height=0.4\textheight,width=\linewidth]{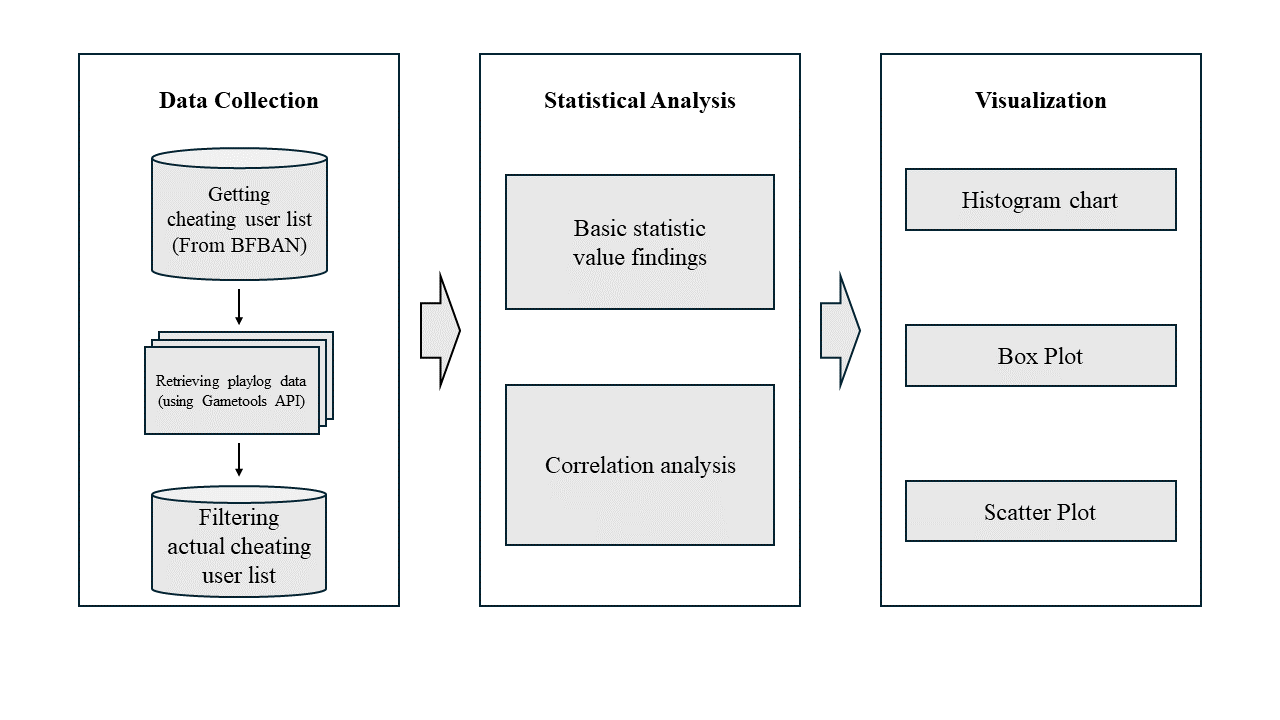}
        \vspace{-1cm}
    \caption{Overall procedure of FPS cheating play detection}
    \label{fig:procedure}
\end{figure}

\vspace{-0.5cm}


Fig.~\ref{fig:procedure} shows the overall procedure of the proposed method. We aimed to conduct an in-depth analysis of cheating players using statistical analysis. In this process, we calculated basic statistics such as mean, standard deviation, and minimum/maximum values for each variable. Then we extracted meaningful variables based on these calculations. This approach allowed us to understand the overall characteristics of the data and identify key factors.

First, to understand the data distribution, we calculated basic statistics for each variable, including mean, standard deviation, minimum value, maximum value, median, and quartiles. This helped us grasp the central tendency and variability of the data and check for outliers or anomalies.

Second, we analyzed the correlations between variables. Then, we identified the relationships between variables through correlation analysis; and we determined which variables have influence value to others. This provided crucial information for understanding the interactions between variables. By measuring correlation coefficients, we identified variables that showed significant correlations.

Finally, for the data visualization, we created various graphs such as histograms, box plots, and scatter plots. Histograms allowed us to visually confirm the distribution of each variable, while box plots provided a clear view of the quartiles, median, and outliers. Scatter plots were used for visually expressing the relationships between two variables, making it easier to understand correlations or trends between variables.

By visually representing the results of complex statistical analyses, we could easily identify key characteristics and patterns in the data. Additionally, using various graphs to illuminate different aspects of the data enabled a more comprehensive and in-depth analysis.

\section{Experiment Result}

\subsection{Dataset}
Due to the absence of publicly available datasets for experimenting with the methodology of this paper, we collected the data directly. The dataset was gathered using the ``Gametools API"~\cite{gametools} with the Requests~\cite{Requests} library. The ``Gametools API" allows us to fetch the list and information of users identified as cheating players from ``BFBAN"~\cite{bfban}. ``BFBAN" is a site where users can report cheating players and check the status of their bans online. We specifically retrieved the list of users banned by the game company and analyzed their statistics.

The data was collected in two stages. First, we collected the list of cheating players from the ``Gametools API". Second, we gathered detailed information about these users' cheating activities using the same API.

The total number of cheating players is 119,569. The data spans from the first reported cheating player on October 27, 2018, to the most recent report on April 29, 2024. However, due to some data being missing during the collection process from the ``Gametools API", we were only able to collect data for 44,307 users.

\subsection{Explanation of Key Variables}
The variables \textit{hack\_score} and \textit{hack\_score\_current} indicate the extent of a user's hacking activities. The \textit{hack\_score} represents the highest value of either \\ \textit{hack\_score\_current} or the player's highest historic \textit{hack\_score}, 
while the \\ \textit{hack\_score\_current} represents the \textit{hack\_score} of the player's current stats. Hack Score measures how abnormal and hacker-like a player's stats are. When the \textit{hack\_score} is more than 100, that means their stats are highly abnormal and can only be produced by hacks (except rare false positives. Refer to the dataset's description; the false positive ratio is 0.5\% in total.). A \textit{hack\_Score} from 0-100 is calculated for each weapon the player has used. The \textit{stats.kills} variable represents the number of opponents a user has eliminated in the game. The \textit{stats.kpm} shows how many opponents the user kills per minute in the game. The \textit{stats.rank} indicates the user's level in the game, with higher values showing that the user has gained more experience points. The \textit{unreleased\_weapons.kills} represents the number of kills a user has made using weapons that have not yet been officially released, indicating kills made with special or test weapons.

\subsection{Basic Statistical Analysis}
The dataset consists of 44,307 observations and includes various variables. Basic statistics, such as mean, standard deviation, minimum, maximum, median, and quartiles, were calculated for key variables (hack\_score, hack\_score\_current, stats.kills, stats.kpm, stats.rank, unreleased\_weapons.kills). These calculations allowed us to understand the central tendency and variability of the data and identify any outliers or anomalies. The whole result of the basic analysis is summarized in Table 1.

\begin{table}[ht]
    \centering
    \begin{adjustbox}{width=\linewidth}
    \begin{tabular}{|c|r|r|r|r|r|r|r|r|}
    \hline
    \textbf{Column} & \textbf{Mean} & \textbf{Std} & \textbf{Min} & \textbf{Q1: 25\%} & \textbf{Q2: 50\%} & \textbf{Q3: 75\%} & \textbf{Max} \\ \hline
    hack\_score & 477.45 & 736.96 & 0 & 41.75 & 143.0 & 528.25 & 16667.0 \\ \hline
    \makecell{hack\_score\\\_current} & 473.89 & 732.86 & 0 & 41.75 & 143.0 & 528.25 & 16667.0 \\ \hline
    stats.kills & 6937.79 & 14906.77 & 0 & 527.25 & 2958.5 & 7826.5 & 781157.0 \\ \hline
    stats.kpm & 1.13 & 6.74 & 0 & 0.33 & 0.64 & 1.02 & 1488.56 \\ \hline
    stats.rank & 65.45 & 46.87 & 0 & 22.0 & 61.0 & 109.0 & 154.0 \\ \hline
    \makecell{unreleased\\\_weapons.\\kills} & 6937.79 & 14906.77 & 0 & 527.25 & 2958.5 & 7826.5 & 781157.0 \\ \hline
    \end{tabular}
    \end{adjustbox}
    \label{tab:summary_stats}
    \caption{Summary Statistics of Key Variables}
\end{table}

\vspace{-1cm}

\subsection{Correlation Analysis Between Variables}
The hack\_score and hack\_score\_current showed a high correlation of 0.996, indicating that users who have cheated once tend to continue cheating. The stats.kills, and stats.rank showed a strong correlation of 0.794, meaning that recording a high number of kills increases the likelihood of reaching higher levels. The hack\_score and stats.rank showed a moderate correlation of 0.449, suggesting that the relationship between cheating and level is not significant. This result indicates that low-level users are not necessarily more prone to cheating and high-level users are not guaranteed to refrain from cheating. The stats.kills and unreleased\_weapons.kills showed a perfect positive correlation, indicating that users who use unreleased weapons (cheating players) have a 100\% probability of recording a high number of kills. This suggests that some cheating players use these cheats to generate high number of kills, adversely affecting innocent opponents. The whole result of the correlation analysis is summarized in Table 2.

\begin{table}[ht]
\centering
\begin{adjustbox}{width=\linewidth}
\begin{tabular}{|c|r|r|r|r|r|r|}
\hline
\textbf{} & \textbf{hack\_score} & \textbf{\makecell{hack\_score\\\_current}} & \textbf{stats.kills} & \textbf{stats.kpm} & \textbf{stats.rank} & \textbf{\makecell{unreleased\\\_weapons.\\kills}} \\ \hline
\textbf{hack\_score} & 1.000000 & 0.996151 & 0.385814 & 0.015135 & 0.449039 & 0.385814 \\ \hline
\textbf{\makecell{hack\_score\\\_current}} & 0.996151 & 1.000000 & 0.379819 & 0.015231 & 0.442342 & 0.379819 \\ \hline
\textbf{stats.kills} & 0.385814 & 0.379819 & 1.000000 & 0.004692 & 0.793550 & 1.000000 \\ \hline
\textbf{stats.kpm} & 0.015135 & 0.015231 & 0.004692 & 1.000000 & -0.006504 & 0.004692 \\ \hline
\textbf{stats.rank} & 0.449039 & 0.442342 & 0.793550 & -0.006504 & 1.000000 & 0.793550 \\ \hline
\textbf{\makecell{unreleased\\\_weapons.\\kills}} & 0.385814 & 0.379819 & 1.000000 & 0.004692 & 0.793550 & 1.000000 \\ \hline
\end{tabular}
\end{adjustbox}
\label{tab:correlation_matrix}
\caption{Correlation Matrix between key features}
\end{table}

\subsection{Data Visualization}
We visualized the result using histograms, box plots, and scatter plots.

\begin{figure}[!ht]
\includegraphics[width=\linewidth, height=0.65\textheight]{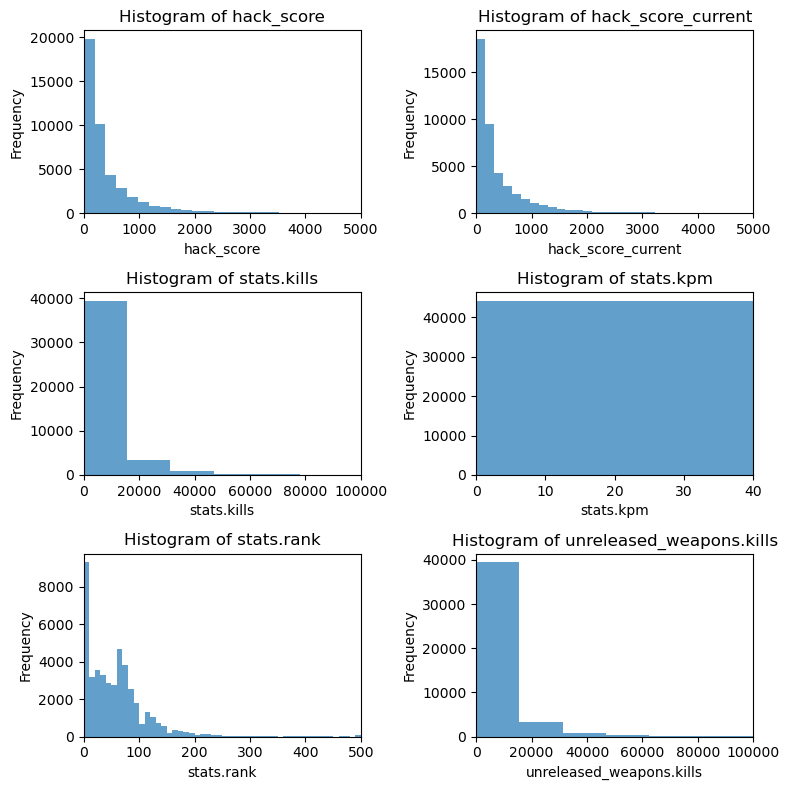}
    \caption{Histogram of key features}
    \label{fig:histogram}
\end{figure}

The histogram chart is used to visually confirm the distribution of variables and identify data concentrated in specific ranges and outliers. The Histogram chart is shown in Fig~\ref{fig:histogram}.

\begin{figure}[!ht]
\includegraphics[width=\linewidth, height=0.60\textheight]{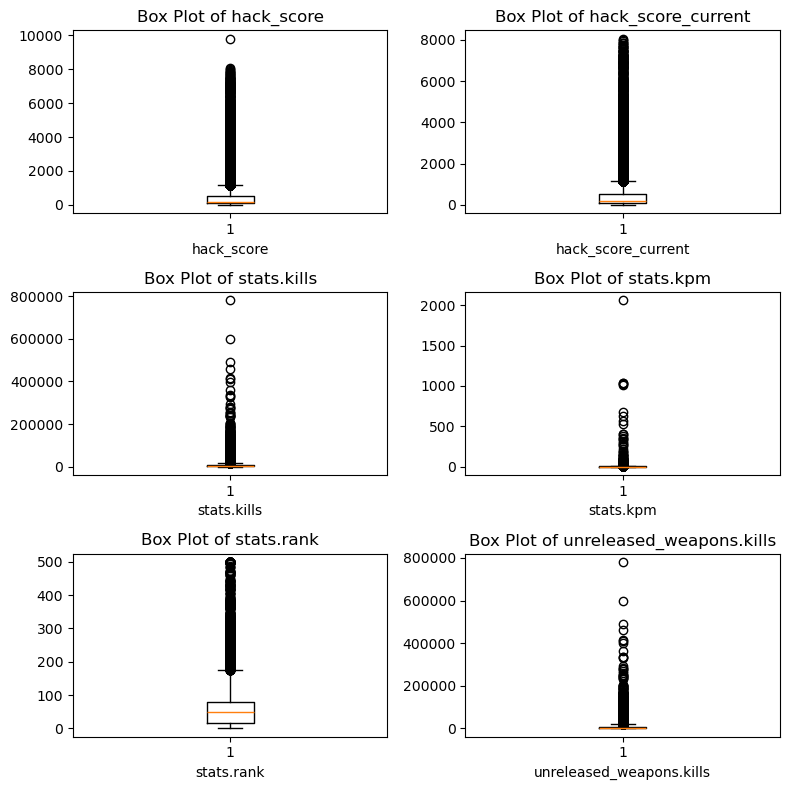}
    \vspace{-0.5cm}
    \caption{Box Plot of key features}
    \vspace{-0.5cm}
    \label{fig:boxplot}
\end{figure}

The box Plot is used to visually analyze the median, quartiles, and outliers of the variables to understand the overall data distribution. The Box Plot result is presented in Fig~\ref{fig:boxplot}.

\begin{figure}[!ht]
\includegraphics[width=\linewidth, height=0.65\textheight]{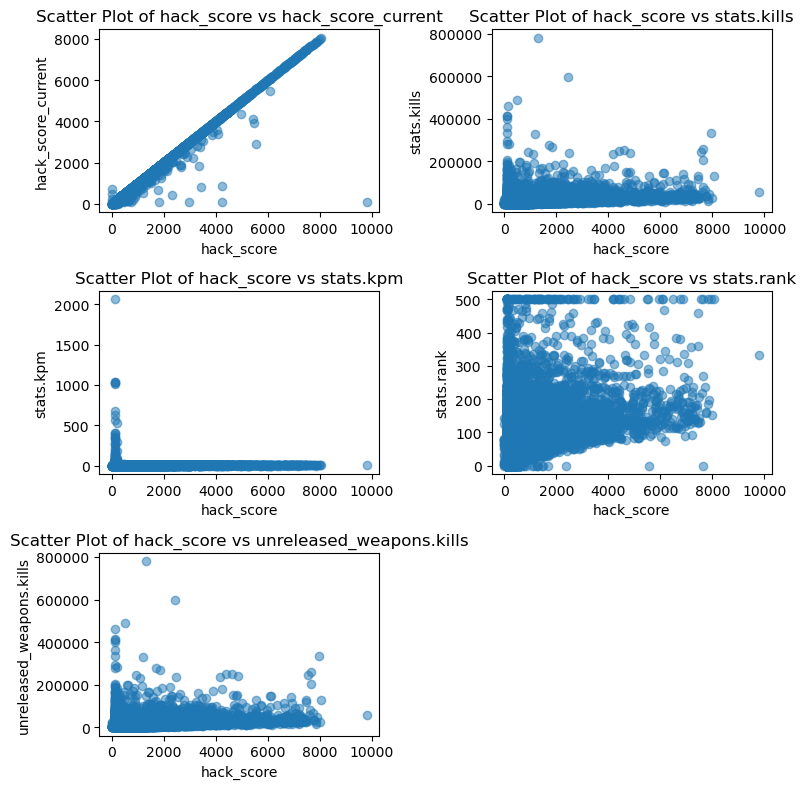}
    \caption{Scatter between hack\_score and other variables}
    \label{fig:Scatter}
\end{figure}

The scatter Plot is used to visually express the relationship between hack score and other variables, making it easy to understand correlations and trends between variables. The Scatter Plot result is presented in Fig~\ref{fig:Scatter}.

The key findings from the visualization process are as follows.
\begin{itemize}
    \item Most benign users have lower hack\_score than the median value. It indicates that most users do not use cheating programs. 
    \item We observe the hack\_score and hack\_score\_current have a similar distribution. That means a user who does a cheating play tends to keep on doing cheating play continuously. 
    \item We find the cheating player's records do not show a higher kill count (stats.kills). On the one hand, it means the cheating program is not always effective for all opponent users (some high-level human players can still overwhelm the cheating players.) On the other hand, it also mean the cheating tool can limit the kill ratio not to be detected by an in-game monitoring system. The lower kpm value (stats.kpm < 40) supports this point of view. Also, some cheating players use unknown weapons to kill the opponent players. 
\end{itemize}

The proposed method does not require a high-performance computing environment. The simple statistical values can be estimated in near real-time by streaming in-game play logs. 
In particular, the continuous cheating behavior of users with high hack scores and the perfect correlation between stats.kills and unreleased\_weapons.kills can serve as strong indicators for identifying cheating players. These findings can be beneficial for detecting and responding to cheating behavior in real-time within the game.

\section{Discussion}

We gathered the dataset from the several public sites. and the banned user list can be regarded as reliable. The game company does not provide the official dataset, including banned users, to the public; thus, our approach is the best way so far. However, the dataset does not include a complete cheating player list (\textit{i.e.}, it has a false-negative error) because it relies on the volunteer users' reporting. Also, the dataset can have a false-positive issue when the reporting is intentionally misreported (to dishonor a specific opponent user).

In this study, we adopt simple methods such as correlation analysis and statistical visualization to detect cheating play. As a result, we identified general patterns of cheating players through the correlation analysis between key variables. In the future, we will explore using more statistical techniques for robust and lightweight cheating play detection. We will consider testing other FPS games to get more generalizations of the proposed method. Also, we will research anomaly detection based on malicious and benign user data to overcome the drawbacks of our methods. Due to the difficulty in obtaining data from normal players, there are limitations in creating an anomaly detection model. We plan to implement an anomaly detection model based on the kill-log of abnormal players and explore this topic further.


\section{Conclusion}
This study proposed a data-driven approach to address the issue of cheating in the Battlefield. To conclude, the main contributions of this study are as follows:
First, we confirm the repeatability of cheating behavior. The high correlation (0.996) between hack\_score and hack\_score\_current indicates that users who start cheating tend to continue cheating. This is a significant finding as it shows that cheating behavior is not a one-time occurrence but follows a repetitive pattern. These results emphasize the need for continuous monitoring and response to cheating players in cheating detection systems.

Second, we discovered that the use of specific weapons could be an indicator of cheating behavior. The perfect correlation (1.000) between stats.kills and unreleased\_weapons.kills indicates that cheating players tend to record many kills using unreleased weapons. This suggests that monitoring the use of specific weapons can be an important indicator for detecting cheating. Game developers can use this pattern to detect cheating behavior more effectively.

Last, we showed that cheating behavior can occur at all player levels. The moderate correlation (0.449) between hack\_score and stats.rank indicates that cheating players are not necessarily concentrated at high levels. This implies that cheating can frequently occur even at lower levels, highlighting the need for cheating detection across all player levels.


\subsubsection{\ackname} 
This work was supported by the Institute for Information \& Communications
Technology Promotion (IITP) grant funded by the Korea government (MSIT).
(Grant No. 2020-0-00374, Development of Security Primitives for Unmanned
Vehicles).

%
%
%
\bibliography{main}

\begin{thebibliography}{10}
\providecommand{\url}[1]{\texttt{#1}}
\providecommand{\urlprefix}{URL }
\providecommand{\doi}[1]{https://doi.org/#1}

\bibitem{gametools}
API, G.: \url{https://api.gametools.network/docs/}, accessed: 2024-08-09

\bibitem{bfban}
BFBAN: \url{https://admin.bfban.com/player}, accessed: 2024-08-09

\bibitem{battlefieldgames}
EA: Battlefield games. \url{https://www.ea.com/ko-kr/games/battlefield}, accessed: 2024-08-09

\bibitem{han2022cheating}
Han, M.L., Kwak, B.I., Kim, H.K.: Cheating and detection method in massively multiplayer online role-playing game: systematic literature review. IEEE Access  \textbf{10},  49050--49063 (2022)

\bibitem{han2015online}
Han, M.L., Park, J.K., Kim, H.K.: Online game bot detection in fps game. In: Proceedings of the 18th Asia Pacific Symposium on Intelligent and Evolutionary Systems-Volume 2. pp. 479--491. Springer (2015)

\bibitem{jonnalagadda2021robust}
Jonnalagadda, A., Frosio, I., Schneider, S., McGuire, M., Kim, J.: Robust vision-based cheat detection in competitive gaming. Proceedings of the ACM on Computer Graphics and Interactive Techniques  \textbf{4}(1),  1--18 (2021)

\bibitem{liu2017detecting}
Liu, D., Gao, X., Zhang, M., Wang, H., Stavrou, A.: Detecting passive cheats in online games via performance-skillfulness inconsistency. In: 2017 47th Annual IEEE/IFIP International Conference on Dependable Systems and Networks (DSN). pp. 615--626. IEEE (2017)

\bibitem{may2024thinking}
May, L., Hall, B.: Thinking ecologically with battlefield 2042. Game Studies  \textbf{24}(1) (2024)

\bibitem{nie2024vadnet}
Nie, B., Ma, B.: Vadnet: Visual-based anti-cheating detection network in fps games. Traitement du Signal  \textbf{41}(1) (2024)

\bibitem{pinto2021deep}
Pinto, J.P., Pimenta, A., Novais, P.: Deep learning and multivariate time series for cheat detection in video games. Machine Learning  \textbf{110}(11),  3037--3057 (2021)

\bibitem{Requests}
Reitz, K.: Requests: Http for humans™. \url{https://requests.readthedocs.io/en/latest/}, accessed: 2024-08-09

\bibitem{tekofsky2015past}
Tekofsky, S., Spronck, P., Goudbeek, M., Plaat, A., van Den~Herik, J.: Past our prime: A study of age and play style development in battlefield 3. IEEE Transactions on Computational Intelligence and AI in Games  \textbf{7}(3),  292--303 (2015)

\bibitem{battlefieldgamesimg}
terDwas: Hurricane loader in the battlefield 2042. \url{https://github.com/terDwas/Battlefield-2042-Hurricane-Cheat}, accessed: 2024-08-09

\end{thebibliography}
\bibliographystyle{splncs04}

\end{document}